\documentclass[a4paper,11pt]{article}
\usepackage[dvips]{graphicx}
\usepackage{amssymb}
\usepackage{amsmath}

\usepackage{cite}

\input arrow

\begin{document}

\title{Extended Supersymmetric Quantum Mechanics Algebras in Scattering States of Fermions off Domain Walls}
\author{
K. Kleidis{\,} and V.K. Oikonomou\thanks{voiko@physics.auth.gr}\\
Department of Mechanical Engineering\\ Technological 
Education Institute of Central Macedonia \\
62124 Serres, Greece \\
} \maketitle

\begin{abstract}
We study the underlying extended supersymmetric structure in a system composed of fermions scattered off an infinitely extended static domain wall in the $xz$-plane. As we shall demonstrate, the fermionic scattered states are associated to two $N=2$ one dimensional supersymmetric quantum mechanical algebras with zero central charge. These two symmetries are combined to form a non-trivial one dimensional $N=4$ superalgebra with various central charges. In addition, we form higher dimensional irreducible representations of the two $N=2$ algebras. Moreover, we study how the Witten index behaves under compact odd and even perturbations, coming from a background magnetic field and some non-renormalizable Yukawa mass terms for the fermions. As we shall demonstrate, the Witten index is invariant only when the magnetic field is taken into account and particularly when only the $z$-component of the field is taken into account. Finally, we study the impact of this supersymmetric structures on the Hilbert space of the fermionic states and also we present a deformed extension of the $N=2$ supersymmetric structure.
\end{abstract}

\section*{Introduction and Motivation}

Extended topological structures such as domain walls, cosmic strings and monopoles, are theoretical predictions of grand unified theories and naturally occur in many theoretical frameworks \cite{vilenkin,lazaridesvasiko}. From the three, the most plausible for cosmological reasons, are cosmic strings associated with high energy symmetry breaking scale. On the contrary, monopoles and domain walls may lead to cosmological inconsistencies in reference to the observed universe. However, in the case when the domain walls are topologically unstable, domain walls are phenomenologically acceptable, since these disappear before dominating the expansion of the universe. Moreover, this kind of domain walls are locally stable and may lose their energy through interactions with the surrounding medium. Theoretical arguments and recent analysis of WMAP data \cite{campanelli}, indicate that the presence of a low tension domain wall network in the universe is not ruled out, providing another somehow natural and non-exotic alternative to existing dark energy models related to modified gravity theories (for an important stream of papers on the latter theories see \cite{odintsovgravity} and references therein). The interaction of the domain wall network with the surrounding plasma in the early universe determines the evolution of domain walls. To this end, the interaction of matter fermions with the domain walls enjoys an elevated role among all other plasma interactions. With respect to the interaction of the fermions with the domain walls, two cases are studied in the literature, namely, the fermion scattering off domain walls \cite{campanelli} and the existence of bound states zero modes near the walls \cite{lazaridesvasiko}. In order to fully understand the evolution of the primordial domain wall network, it is necessary to understand the interaction of the domain walls with primordial matter fermions, among other interactions, for example domain wall interactions with scalar fields. And this is owing to the fact that fermionic fields like the neutrinos, may have some imprints of this primordial interaction on their energy spectrum. Moreover, in reference to neutrinos, since the nature of the neutrino (Dirac or Majorana) is yet to be understood-revealed, studies of such interactions may provide us with important information. 

Supersymmetry is one of the most used tools in the beyond the Standard Model quantum field theory physics and also plays an important role in most modern sting theory models. Although there is no experimental verification of supersymmetry, the theoretical and phenomenological attributes of this graded super-Poincare algebra are so useful in model building, that rendered supersymmetry one necessary ingredient of various fields of research. In reference to experimental verification, supersymmetry is obviously broken in our world, and to this end there exist various ways to break supersymmetry. For an important stream of papers in reference to supersymmetry breaking in field theoretic grand unified theories see \cite{odi1,odi2} and for some cosmological and supergravity theories applications see \cite{odi3}. 

Supersymmetric quantum mechanics (abbreviated to SUSY QM hereafter), was firstly introduced to model supersymmetry breaking in quantum field theory \cite{witten1}. In time, SUSY QM has developed to be a powerful tool for integrability and dimensionally reduced quantum field theories. For important reviews and textbooks on SUSY QM, see \cite{reviewsusyqm} and references therein. Nowadays, SUSY QM is an independent research field, with numerous applications in various research areas. Particularly, mathematical aspects of Hilbert spaces corresponding to SUSY QM systems and also applications to various quantum mechanical systems were performed in \cite{diffgeomsusyduyalities} and \cite{various,susyqminquantumsystems,plu1,plu2}. In addition, extended supersymmetries and harmonic superspaces or gravity, are presented in \cite{extendedsusy,ivanov} and applications of SUSY QM to scattering appear in \cite{susyqmscatter}. Particular features of supersymmetry breaking were studied in \cite{susybreaking}. Aspects of the possible connection between central charge extended SUSY QM and global four dimensional spacetime supersymmetry were presented in \cite{ivanov}.  

In this paper the focus is on scattered fermions off domain walls, and particularly on a specific attribute that the system of scattered fermions has. As we shall demonstrate, the scattered fermionic states constitute the Hilbert space of two independent $N=2$, $d=1$ supersymmetries. These supersymmetries can be combined to form an $N=4$, $d=1$ SUSY QM algebra with central charge. The latter result holds generally and not under some specific assumptions. The existence of non-trivial SUSY QM algebras in fermionic systems around defects is potentially interesting and we shall present all the details to prove that SUSY QM is unbroken in all cases. Having in mind that since the scattering of fermions off domain walls together with the presence of localized fermionic zero modes could potentially influence the evolution of domain walls in the early universe, the existence of an one dimensional supersymmetry in the fermionic system is a rather intriguing result. Particularly the fact that the supersymmetric structure is highly non-trivial due to the existence of the extended $N=4$ SUSY QM algebra. 
 
This paper is organized as follows: In section 1 we present the essentials of fermionic states scattering off domain walls. Particularly we shall be interested in the scattering solutions of the fermionic equations of motion, which will be the starting point of our analysis. In section 2, we shall demonstrate that two independent $N=2$, $d=1$ SUSY QM algebras underlie the fermionic system and also present some higher dimensional reducible representations of these two SUSY QM algebras. Moreover, a spin complex structure can be naturally defined using the Hilbert spaces of the fermionic states. In the end of section 2, we shall trivially extend one of these two algebras to be a $q$-deformed SUSY QM algebra. In addition, we study the effect of even and odd compact perturbations of the Witten index. In section 3, we shall present that these two $N=2$, $d=1$ algebras can be combined to form a central charge extended $N=4$ SUSY QM algebra. In section 4, we shall present the implications of SUSY QM on the Hilbert space of the scattered fermionic states. Particularly, we shall describe the local geometrical implications of the SUSY QM algebra and also we shall demonstrate that there exists an underlying product of independent global $U(1)$ symmetries corresponding to the scattered fermionic states. The conclusions follow in the end of the article.

\section{Preliminaries}

In this section we briefly describe the theoretical framework which we shall make use of. We shall follow the notation and results of reference \cite{campanelli}. A domain wall is created when a discrete symmetry of a quantum grand unified gauged system is spontaneously broken, with the discrete broken symmetry not being part of the gauge symmetry of the total gauge theory. As a result, the total vacuum manifold contains various distinct vacuum states, with only one of them being the true vacuum state of the field related to the spontaneous symmetry breaking of the discrete symmetry \cite{lazaridesvasiko}. Domain walls create a natural spatial separation of these different vacuum regions and as a result, the symmetry breaking related field interpolates between these distinct vacuum states. Following \cite{campanelli}, we shall consider a static domain wall in the $xz$-plane. The Lagrangian of the model under study is equal to: 
\begin{align}\label{Lagranena}
&\mathcal{L}=\frac{1}{2}\partial_{\mu}\Phi\partial^{\mu}\Phi-\frac{\lambda}{4}(\Phi^2-\eta^2)^2+i\bar{\psi}_L\gamma^{\mu}\partial_{\mu}\psi_L
+i\bar{\psi}_R\gamma^{\mu}\partial_{\mu}\psi_R \\ \notag &-(g_D\Phi\bar{\psi}_L\psi_R+g_L\Phi\bar{\psi}_L\psi_L^c+g_R\Phi\bar{\psi}_R\psi_R^c+\mathrm{h.c})
\end{align}
The scalar field $\Phi$ will describe the domain wall and its vacuum states, between which the field interpolates, are equal to:
\begin{equation}\label{interpol}
\langle \Phi \rangle =\eta,{\,}{\,}{\,}\langle \Phi \rangle =-\eta
\end{equation}
The region where the field is out of these vacuum states, exactly describes the domain wall, which is equal to:
\begin{equation}\label{domwallsol}
\Phi(y) =\eta \tanh (\frac{y}{\Delta_1}),
\end{equation}
with $\Delta_1 =\sqrt{2\lambda}/\eta$, the domain wall thickness. Notice that the Lagrangian (\ref{Lagranena}) contains both Dirac and Majorana Yukawa terms, with couplings $g_D$ and $g_L,g_R$ respectively. In addition, $\psi^c=C\bar{\psi}^T$, with $C=i\gamma^2\gamma^0$ the charge conjugation operator. Now we shall demonstrate that there exist scattering states in the system. For details see \cite{campanelli}. In the broken phase $y\rightarrow \pm \infty$, the field $\Phi$ takes the values $\pm \eta$, and the Dirac equation looks like:  
\begin{equation}\label{fde}
(i\gamma^{\mu}\partial_{\mu}-G\Phi)\Psi=0
\end{equation}
where,
\begin{equation}\label{fhfhfhf}
G=\bigg{(}\begin{array}{ccc}
 0 & \mathcal{G} \\
  \mathcal{G} & 0  \\
\end{array}\bigg{)},{\,}{\,}{\,}\mathcal{G}=\bigg{(}\begin{array}{ccc}
 g_L & g_D \\
  g_D & g_R  \\
\end{array}\bigg{)},{\,}{\,}{\,}\Psi=\left ( \begin{array}{c}
 \psi_L \\
  \psi_R^c \\
  \psi_L^c \\
  \psi_R \\
\end{array}\right )
\end{equation}
For $y\rightarrow \pm \infty$, the Dirac equation becomes:
\begin{equation}\label{sde}
(i\gamma^{\mu}\partial_{\mu}\mp M)\Psi=0,{\,}\mathrm{if}{\,}y\rightarrow \pm \infty
\end{equation}
with $M=\eta G$. Diagonalizing $M$, we get:
\begin{equation}\label{sde1}
(i\gamma^{\mu}\partial_{\mu}\mp \Delta)\Psi_M=0,{\,}\mathrm{if}{\,}y\rightarrow \pm \infty
\end{equation}
with the matrix $\Delta$ being equal to:
\begin{equation}\label{edffa}
\Delta= \left ( \begin{array}{cccc}
  m_1 & 0 & 0 & 0 \\
  0 & m_2 & 0 & 0 \\
0 & 0 & -m_1 & 0 \\
0 & 0 & 0 & -m_2  \\
\end{array} \right)
\end{equation}
In the above, it is supposed that there exists a matrix $U$, that diagonalizes $M$, so that $\Delta =U^{\dag}MU$. In addition we have $\Psi_M=U^{\dag}\Psi$. Moreover the parameters $m_{1,2}$ are given by the following relation:
\begin{equation}\label{pamm1m2}
m_{1,2}=\frac{1}{2}\Big{[}m_L+m_R\pm\sqrt{4m_D^2+(m_L-m_R)^2}\Big{]},{\,}{\,}{\,}m_D=g_D\eta,{\,}m_L=g_L\eta,{\,}m_R=g_R\eta
\end{equation}
Note that the massive fermion states are Majorana particles. For simplicity we define the fermionic states $\psi_L(y)$ and $\psi_R(y)$ as follows: 
\begin{equation}\label{psifunctions}
\psi_L(y,t)=\left ( \begin{array}{c}
  \alpha(y,t) \\
  \beta (y,t)\\
-\alpha (y,t)\\
-\beta (y,t)\\
\end{array} \right),{\,}{\,}{\,}\psi_R(y,t)=\left ( \begin{array}{c}
  \gamma (y,t)\\
  \delta (y,t)\\
\gamma (y,t)\\
\delta (y,t)\\
\end{array} \right)
\end{equation}
The equations of motion (\ref{sde1}) can be written as follows:
\begin{align}\label{eqnmotionsdffdsenewlife}
& \Phi''-\lambda\Phi(\Phi^2-\eta^2)=4g_D\mathrm{Re}(a^*\gamma-\beta^*\delta)\\ \notag &
\beta'+i\dot{\alpha}=g_D\Phi\gamma+g_L\Phi\beta^*\\ \notag &
\alpha'-i\dot{\beta}=g_D\Phi\delta+g_L\Phi\alpha^*\\ \notag &
\delta'+i\dot{\gamma}=g_D\Phi\alpha+g_R\Phi\delta^*\\ \notag &
\gamma'-i\dot{\delta}=g_D\Phi\beta+g_R\Phi\alpha^*\\ \notag &
\end{align}
Focusing on solutions for which the back-reaction of the fermionic field on the domain wall is null, and making the ansatz $\beta=\alpha ^*$ and $\gamma =\delta^*$, we may seek solutions of the form:
\begin{equation}\label{ggfg}
a(y,t)=a_+(y)e^{-iEt}+a_(y)e^{iEt},{\,}{\,}{\,}\delta (y,t)=\delta_+(y)e^{-iEt}+\delta_{-}e^{iEt}
\end{equation}
with $E$ the energy of the particle. Then, for $y>0$ and in the thin-wall approximation ($\Delta_1 \rightarrow 0$), we get the following solutions:
\begin{align}\label{fgderdr}
& a_{+}=c_1e^{ip_1y}+c_2e^{ip_2y}+c_3e^{-ip_1y}+c_4e^{-ip_2y}, \\ \notag &
a_{-}^*=ix_1c_1e^{ip_1y}+ix_2c_2e^{ip_2y}-ix_1c_3e^{-ip_1y}-ix_1c_4e^{-ip_2y}, \\ \notag &
\delta_{+}=ix_3c_1e^{ip_1y}+ix_4c_2e^{ip_2y}-ix_3c_3e^{-ip_1y}-ix_4c_4e^{-ip_2y}, \\ \notag &
\delta_{-}^*=x_5c_1e^{ip_1y}+x_6c_2e^{ip_2y}+x_5c_3e^{-ip_1y}+x_6c_4e^{-ip_2y}, \\ \notag &
\end{align}
with $c_i$ integration constants, and the parameters $p_{1,2}$ and $x_{1},x_2,...,x_6$, which depend on $E$ and $m_{1,2},m_{D,R}$ can be found in \cite{campanelli}.
Then, decomposing the fermionic fields $\psi_{L,R}$ as follows,
\begin{equation}\label{hedd}
\psi_L=\psi_L^{(+)}+\psi_{L}^{(-)},{\,}{\,}{\,}{\,}\psi_{R}=\psi_{R}^{(+)}+\psi_R^{(-)}
\end{equation}
the fermionic solutions can be written in the form:
\begin{equation}\label{psifunctions}
\psi_L^{(\pm)}=\left ( \begin{array}{c}
  \alpha_{\pm} \\
  \alpha_{\mp}^* \\
-\alpha_{\pm} \\
-\alpha_{\mp}^* \\
\end{array} \right)e^{\mp iEt},{\,}{\,}{\,}\psi_R^{(\pm)}=\left ( \begin{array}{c}
  \delta_{\mp}^* \\
  \delta_{\pm} \\
\delta_{\mp}^* \\
\delta_{\pm} \\
\end{array} \right)e^{\mp iEt}
\end{equation}
Similar solutions exist for $y<0$. What we shall need mostly from this section is the existence of scattering states for some energy $E$ of the fermionic particle and as can be seen from relation (\ref{fgderdr}), there exists one such solution for each energy $E$. We shall make extensive use of this solution existence feature of the system in the following sections. In reference to the energies, let us add that the thin-wall approximation is valid whenever the wavelength of the scattered particles is much greater that $\Delta_1$, that is, the thickness of the wall.

\section{$N=2$ Supersymmetric Quantum Mechanics Algebras in the Fermionic Scattering States}

\subsection{Scattered Fermionic States and $N=2$ SUSY QM}

Having established the result that, for each energy of the scattered particle, there exists a scattering solution of the form (\ref{fgderdr}) and (\ref{psifunctions}), we rewrite the Dirac equations of motion (\ref{sde1}) in the following form:
\begin{equation}\label{actualeqns}
\bigg{(}\begin{array}{cc}
 i\sigma^3\partial_0-\sigma^2\partial_1+\sigma^1\partial_2 & i\partial_3 \\
  -i\partial_3 & -i\sigma^3\partial_0+\sigma^2\partial_1-\sigma^1\partial_2  \\
\end{array}\bigg{)}\Psi_M\mp \bigg{(}\begin{array}{ccc}
 \mathcal{M} & 0 \\
  0 & -\mathcal{M}  \\
\end{array}\bigg{)}\Psi_M=0
\end{equation}
with,
\begin{equation}\label{actumaleqns}
\mathcal{M}=\bigg{(}\begin{array}{ccc}
 m_1 & 0 \\
  0 & m_2  \\
\end{array}\bigg{)}
\end{equation}
and $m_{1,2}$ defined in relation (\ref{pamm1m2}). In addition,
\begin{equation}\label{djfjjeerd}
\Psi_M=\left ( \begin{array}{c}
\Psi_A\\
  \Psi_B \\
\end{array}\right )
\end{equation}
with,
\begin{equation}\label{djfjjeerd}
\Psi_A=\left ( \begin{array}{c}
\psi_L\\
  \psi_R^c \\
\end{array}\right ),{\,}{\,}{\,}\Psi_B=\left ( \begin{array}{c}
\psi_L^c\\
  \psi_R \\
\end{array}\right )
\end{equation}
In equation (\ref{actualeqns}), the operators $\sigma_i$, with $i=1,2,3$, denote the Pauli matrices. These equations can be cast in the following forms:
\begin{align}\label{finaleqns}
&(i\sigma^3\partial_0-\sigma^2\partial_1+\sigma^1\partial_2+\mathcal{M})\Psi_A+i\partial_3\Psi_B=0\\ \notag &
(i\sigma^3\partial_0-\sigma^2\partial_1+\sigma^1\partial_2-\mathcal{M})\Psi_B+i\partial_3\Psi_A=0
\end{align}
which is the first set of equations appearing in (\ref{actualeqns}), and also the other set of equations take the form:
\begin{align}\label{finaleqns1}
&(i\sigma^3\partial_0-\sigma^2\partial_1+\sigma^1\partial_2-\mathcal{M})\Psi_A+i\partial_3\Psi_B=0\\ \notag &
(i\sigma^3\partial_0-\sigma^2\partial_1+\sigma^1\partial_2+\mathcal{M})\Psi_B+i\partial_3\Psi_A=0
\end{align}
These two equations shall be the starting point of our analysis, since two $N=2$, $d=1$ supersymmetries can be associated with these two equations. In order to see this, consider the first set of equations, that is, those of relation (\ref{finaleqns}). These equations can be written in terms of the differential operator $\mathcal{D}_1$ as follows:
\begin{equation}\label{transf}
\mathcal{D}_{1}|\Psi_{1}\rangle=0
\end{equation}
with the operator $\mathcal{D}_1$ being equal to:
\begin{equation}\label{susyqmrn567m}
\mathcal{D}_{1}=\left(%
\begin{array}{cc}
i\sigma^3\partial_0-\sigma^2\partial_1+\sigma^1\partial_2+\mathcal{M} & i\partial_3
 \\ i\partial_3 & i\sigma^3\partial_0-\sigma^2\partial_1+\sigma^1\partial_2-\mathcal{M}\\
\end{array}%
\right)
\end{equation}
and the vector $\Psi_1$ equal to:
\begin{equation}\label{ait34e1}
|\Psi_{1}\rangle =\left(%
\begin{array}{c}
  \Psi_A \\
  \Psi_B \\
\end{array}%
\right).
\end{equation}
Equation (\ref{finaleqns}) has solutions which are the scattering states of the previous section, namely (\ref{fgderdr}) and (\ref{psifunctions}). These solutions are the zero mode eigenfunctions of the operator $\mathcal{D}_{1}$. Focusing on the incoming scattering states, there is only one continuous solution $\Psi_1$ which is expressed in terms of the fermionic fields $(\Psi_A,\Psi_B)$, as can be seen from equation and (\ref{fgderdr}) and (\ref{psifunctions}). Thereby, the dimensionality of the kernel of the differential operator $\mathcal{D}_1$ is:
\begin{equation}\label{dimeker}
\mathrm{dim}{\,}\mathrm{ker}\mathcal{D}_{1}=1
\end{equation}
Moreover, in reference to the adjoint of the operator $\mathcal{D}_{1}$, that is $\mathcal{D}_{1}^{\dag}$, 
it is easily verified that:
\begin{equation}\label{dimeke1r11}
\mathrm{dim}{\,}\mathrm{ker}\mathcal{D}_{1}^{\dag}=0
\end{equation}

\noindent The fermionic system of the fields $(\Psi_A,\Psi_B)$ scattered of the domain wall constitutes an unbroken $N=2$, $d=1$ supersymmetry with supercharges and quantum Hamiltonian:
\begin{equation}\label{s7}
\mathcal{Q}_{1}=\bigg{(}\begin{array}{ccc}
  0 & \mathcal{D}_{1} \\
  0 & 0  \\
\end{array}\bigg{)},{\,}{\,}{\,}\mathcal{Q}^{\dag}_{1}=\bigg{(}\begin{array}{ccc}
  0 & 0 \\
  \mathcal{D}_{1}^{\dag} & 0  \\
\end{array}\bigg{)},{\,}{\,}{\,}\mathcal{H}_{1}=\bigg{(}\begin{array}{ccc}
 \mathcal{D}_{1}\mathcal{D}_{1}^{\dag} & 0 \\
  0 & \mathcal{D}_{1}^{\dag}\mathcal{D}_{1}  \\
\end{array}\bigg{)}
\end{equation}
The operators $\mathcal{Q}_{1},\mathcal{Q}^{\dag}_{1},\mathcal{H}_{1}$ satisfy the $N=2$ one dimensional SUSY QM algebra:
\begin{equation}\label{relationsforsusy}
\{\mathcal{Q}_{1},\mathcal{Q}^{\dag}_{1}\}=\mathcal{H}_{1}{\,}{\,},\mathcal{Q}_{1}^2=0,{\,}{\,}{\mathcal{Q}_{1}^{\dag}}^2=0
\end{equation}
In addition, there exists an invariant operator of the system, that commutes with the Hamiltonian, the so called Witten parity, $W$. In our case, this is defined to be:
\begin{equation}\label{wittndrf}
\mathcal{W}=\bigg{(}\begin{array}{ccc}
  1 & 0 \\
  0 & -1  \\
\end{array}\bigg{)}
\end{equation}
This operator satisfies the following relations,
\begin{equation}\label{s45}
[\mathcal{W},\mathcal{H}_{1}]=0,{\,}{\,}{\,}\{\mathcal{W},\mathcal{Q}_{1}\}=\{\mathcal{W},\mathcal{Q}_{1}^{\dag}\}=0,{\,}{\,}{\,}\mathcal{W}^{2}=1
\end{equation}
The Witten parity is an involution operator which renders the quantum Hilbert space of the fermionic system under study $\mathcal{H}$, a $Z_2$-graded space. The $Z_2$ equivalent subspaces of the total Hilbert space of the quantum system is written as:
\begin{equation}\label{fgjhil}
\mathcal{H}=\mathcal{H}^+\oplus \mathcal{H}^-
\end{equation}
The $Z_2$ grading classifies vectors that belong to the two subspaces $\mathcal{H}^{\pm}$, to parity even and parity odd states:
\begin{equation}\label{shoes}
\mathcal{H}^{\pm}=\mathcal{P}^{\pm}\mathcal{H}=\{|\psi\rangle :
\mathcal{W}|\psi\rangle=\pm |\psi\rangle \}
\end{equation}
In addition, the $Z_2$ graded spaces Hamiltonians are:
\begin{equation}\label{h1}
{\mathcal{H}}_{+}=\mathcal{D}_{1}{\,}\mathcal{D}_{1}^{\dag},{\,}{\,}{\,}{\,}{\,}{\,}{\,}{\mathcal{H}}_{-}=\mathcal{D}_{1}^{\dag}{\,}\mathcal{D}_{1}
\end{equation}
The operator $\mathcal{P}^{\pm}$, has eigenstates which we denote $|\psi^{\pm}\rangle$ and call them positive and negative parity eigenstates. These satisfy:
\begin{equation}\label{fd1}
P^{\pm}|\psi^{\pm}\rangle =\pm |\psi^{\pm}\rangle
\end{equation}
Making use of the Witten parity representation (\ref{wittndrf}), the parity eigenstates can be written as follows,
\begin{equation}\label{phi5}
|\psi^{+}\rangle =\left(%
\begin{array}{c}
  |\phi^{+}\rangle \\
  0 \\
\end{array}%
\right),{\,}{\,}{\,}
|\psi^{-}\rangle =\left(%
\begin{array}{c}
  0 \\
  |\phi^{-}\rangle \\
\end{array}%
\right)
\end{equation}
with $|\phi^{\pm}\rangle$ $\epsilon$ $\mathcal{H}^{\pm}$. A criterion that determines whether supersymmetry is broken or not, is the Witten index, which for Fredholm operators, is:
\begin{equation}\label{phil}
\Delta =n_{-}-n_{+}
\end{equation}
with $n_{\pm}$ the finitely many number of zero modes of ${\mathcal{H}}_{\pm}$ in the subspace $\mathcal{H}^{\pm}$. Supersymmetry is certainly unbroken if the Witten index is a non-zero integer, that is $\Delta \neq 0$. In the case the Witten index is zero, that is $\Delta =0$, and if $n_{+}=n_{-}\neq 0$, supersymmetry is unbroken too \cite{reviewsusyqm}. In the case the Witten index is zero and at the same time $n_{+}=n_{-}=0$, supersymmetry is broken.

\noindent The Fredholm index of the operator $\mathcal{D}_{1}$, is directly related to the Witten index of the fermionic system under study, as follows:
\begin{align}\label{ker1}
&\Delta=\mathrm{dim}{\,}\mathrm{ker}
{\mathcal{H}}_{-}-\mathrm{dim}{\,}\mathrm{ker} {\mathcal{H}}_{+}=
\mathrm{dim}{\,}\mathrm{ker}\mathcal{D}_{1}^{\dag}\mathcal{D}_{1}-\mathrm{dim}{\,}\mathrm{ker}\mathcal{D}_{1}\mathcal{D}_{1}^{\dag}=
\\ \notag & \mathrm{ind} \mathcal{D}_{1} = \mathrm{dim}{\,}\mathrm{ker}
\mathcal{D}_{1}-\mathrm{dim}{\,}\mathrm{ker} \mathcal{D}_{1}^{\dag}
\end{align}
Hence, making use of equations (\ref{dimeker}) and (\ref{dimeke1r11}), we can compute the Witten index:
\begin{equation}\label{fnwitten}
\Delta =1
\end{equation}
In conclusion, there is an unbroken $N=2$, $d=1$ supersymmetry underlying the fermionic system of the scattered fermions off the domain wall $(\Psi_A,\Psi_B)$. 

Having computed the Witten index and the Fredholm index of the supersymmetric fermionic system, we now present another attribute of the fermionic system which is due to the SUSY QM algebra and the correspondent $Z_2$ grading. This grading is provided from the Witten parity operator $\mathcal{W}$ to the Hilbert space of the SUSY QM mechanics algebra. The Hilbert space of the scattered fermionic system is divided as follows, 
\begin{equation}\label{jfhrcon}
\mathcal{H}(M)=\mathcal{H}^+(M)\oplus \mathcal{H}^-(M)
\end{equation}
with $M$ denoting spacetime manifold on which the fermions are defined. The $Z_2$ grading provides the fermionic system with a spin complex structure. In order to see this, recall the vectors $|\psi^{+}\rangle$ $\mathcal{H}^+(M)$ and $|\psi^{-}\rangle$ $\in$ $\mathcal{H}^-(M)$, which are equal to, 
\begin{align}\label{phi5gdfghfdh}
&|\psi^{+}\rangle =\left(%
\begin{array}{c}
  |\phi ^{+} \rangle \\
  0 \\
\end{array}%
\right),{\,}{\,}{\,}\in{\,}{\,}\mathcal{H}^+(M)
\\ \notag &
|\psi^{-}\rangle =\left(%
\begin{array}{c}
  0 \\
  |\phi ^{-} \rangle \\
\end{array}%
\right),{\,}{\,}{\,}\in{\,}{\,}\mathcal{H}^-(M),
\end{align}
with $|\phi^{\pm}\rangle$, the vectors corresponding to the operator $\mathcal{D}_1$, defined previously. Recall the form of the supercharges ${\mathcal{Q}}_1$, ${\mathcal{Q}}^{\dag}_1$ of the SUSY QM algebra,
\begin{equation}\label{wit2dyjtyjtyr1}
{\mathcal{Q}_1}=\bigg{(}\begin{array}{ccc}
  0 & {\mathcal{D}}_{1} \\
  0 & 0  \\
\end{array}\bigg{)}, {\,}{\,}{\,}{\,}{\,}{\mathcal{Q}}^{\dag}_1=\bigg{(}\begin{array}{ccc}
  0 & 0 \\
  {\mathcal{D}}_{1}^{\dag} & 0  \\
\end{array}\bigg{)}.
\end{equation}
It worths to see in detail how the supercharges act on the vectors $|\psi^{\pm}\rangle$:
\begin{align}\label{wittyuhjhjhhtyi2}
&\bigg{(}\begin{array}{ccc}
  0 & \mathcal{D}_1 \\
  0 & 0  \\
\end{array}\bigg{)}\left(%
\begin{array}{c}
   0\\
  |\phi ^{-} \rangle \\
\end{array}\right)=\left(%
\begin{array}{c}
  |\phi ^{-} \rangle \\
  0 \\
\end{array}\right),{\,}{\,}{\,}\in{\,}{\,}\mathcal{H}^+(M)
\\ \notag & \bigg{(}\begin{array}{ccc}
  0 & 0 \\
  \mathcal{D}_1^{\dag} & 0  \\
\end{array}\bigg{)}\left(%
\begin{array}{c}
  |\phi ^{+} \rangle \\
  0 \\
\end{array}\right)=\left(%
\begin{array}{c}
   0\\
  |\phi ^{+} \rangle \\
\end{array}\right),{\,}{\,}{\,}\in{\,}{\,}\mathcal{H}^-(M)
.\end{align}
Hence, from the way the supercharges act on the vectors, we can define the following two maps:
\begin{align}\label{mapsscharge}
&{\mathcal{Q}_1}:\mathcal{H}^-(M)\rightarrow \mathcal{H}^+(M)
\\ \notag & {\mathcal{Q}}^{\dag}_1:\mathcal{H}^+(M)\rightarrow \mathcal{H}^-(M)
.\end{align}
These two maps in turn, constitute a two term spin complex which is of the form:
$$\harrowlength=40pt \varrowlength=.618\harrowlength
\sarrowlength=\harrowlength
\mathcal{H}^+(M)\commdiag{\mapright^{{\mathcal{Q}}^{\dag}_1} \cr \mapleft_{{\mathcal{Q}_1}}}\mathcal{H}^-(M)$$
In reference to the Witten index of the SUSY QM fermionic system and the Fredholm index of the operator $\mathcal{D}_1$ the index of this two term spin complex, is equal to the Fredholm index of the operator $\mathcal{D}_1$.

\subsection{A Second $N=2$, $d=1$ Supersymmetry in the Fermionic System}

So far we focused on the set of equations (\ref{finaleqns}) and we associated an $N=2$, $d=1$ SUSY QM algebra with it. The same applies to the second set of equations, namely (\ref{finaleqns1}), that is, another unbroken $N=2$, $d=1$ SUSY QM can be associated with the set of equations (\ref{finaleqns1}). Indeed, these equations can be written in terms of the differential operator $\mathcal{D}_2$ as follows:
\begin{equation}\label{transf}
\mathcal{D}_{2}|\Psi_{2}\rangle=0
\end{equation}
with the operator $\mathcal{D}_2$ being equal to:
\begin{equation}\label{susyqmrn567m1we3}
\mathcal{D}_{2}=\left(%
\begin{array}{cc}
i\sigma^3\partial_0-\sigma^2\partial_1+\sigma^1\partial_2+\mathcal{M} & i\partial_3
 \\ i\partial_3 & i\sigma^3\partial_0-\sigma^2\partial_1+\sigma^1\partial_2-\mathcal{M}\\
\end{array}%
\right)
\end{equation}
and the vector $\Psi_2$ equal to:
\begin{equation}\label{ait34e1}
|\Psi_{2}\rangle =\left(%
\begin{array}{c}
  \Psi_{B} \\
  \Psi_A \\
\end{array}%
\right).
\end{equation}
In this case too, equation (\ref{finaleqns1}) has for solutions the scattering states of the previous section and these solutions are the zero mode eigenfunctions of the operator $\mathcal{D}_{2}$. Following the same line of argument as in the previous case studied in this section, the dimensionality of the kernel of the differential operator $\mathcal{D}_2$ is:
\begin{equation}\label{dimeker}
\mathrm{dim}{\,}\mathrm{ker}\mathcal{D}_{2}=1
\end{equation}
The fermionic system of the fields $(\Psi_B,\Psi_A)$ scattered of the domain wall constitutes an unbroken $N=2$, $d=1$ supersymmetry with supercharges and quantum Hamiltonian:
\begin{equation}\label{sdff7}
\mathcal{Q}_{2}=\bigg{(}\begin{array}{ccc}
  0 & \mathcal{D}_{2} \\
  0 & 0  \\
\end{array}\bigg{)},{\,}{\,}{\,}\mathcal{Q}^{\dag}_{2}=\bigg{(}\begin{array}{ccc}
  0 & 0 \\
  \mathcal{D}_{2}^{\dag} & 0  \\
\end{array}\bigg{)},{\,}{\,}{\,}\mathcal{H}_{2}=\bigg{(}\begin{array}{ccc}
 \mathcal{D}_{2}\mathcal{D}_{2}^{\dag} & 0 \\
  0 & \mathcal{D}_{2}^{\dag}\mathcal{D}_{2}  \\
\end{array}\bigg{)}
\end{equation}
Omitting the details for brevity, the Witten index can be calculated in this case too, and using the same reasoning we may conclude that supersymmetry is unbroken.

\subsection{Extended Supersymmetric-Higher Representation Algebras}

As we have demonstrated, there are two unbroken $N=2$ SUSY QM algebras underlying equations (\ref{finaleqns}) and (\ref{finaleqns1}). A natural question that springs to mind is whether these two supersymmetries can be combined further, to form higher order irreducible representations. As we shall now demonstrate, the two supersymmetries can be combined to form a higher representation of a single $N=2$, $d=1$ supersymmetry. Denoting the supercharges of the higher representation of the $N=2$ SUSY algebra ${\mathcal{Q}}_{H}$ and  ${\mathcal{Q}}_{H}^{\dag}$, these can be written in terms of the operators $\mathcal{D}_1$ and $\mathcal{D}_2$. Indeed the supercharges are equal:
\begin{equation}\label{connectirtyrtons}
{\mathcal{Q}}_{H}= \left ( \begin{array}{cccc}
  0 & 0 & 0 & 0 \\
  {\mathcal{D}}_{1} & 0 & 0 & 0 \\
0 & 0 & 0 & 0 \\
0 & 0 & {\mathcal{D}}_{2}^{\dag} & 0  \\
\end{array} \right),{\,}{\,}{\,}{\,}{\mathcal{Q}}_{T}^{\dag}= \left ( \begin{array}{cccc}
  0 &  {\mathcal{D}}_{1}^{\dag} & 0 & 0 \\
  0 & 0 & 0 & 0 \\
0 & 0 & 0 & {\mathcal{D}}_{2} \\
0 & 0 & 0 & 0  \\
\end{array} \right)
.\end{equation}
Moreover, the Hamiltonian of the combined quantum system $H_H$, is equal to,
\begin{equation}\label{connections1dtr}
H_H= \left ( \begin{array}{cccc}
  {\mathcal{D}}_{1}^{\dag}{\mathcal{D}}_{1} & 0 & 0 & 0 \\
  0 & {\mathcal{D}}_{1}{\mathcal{D}}_{1}^{\dag} & 0 & 0 \\
0 & 0 & {\mathcal{D}}_{2}{\mathcal{D}}_{2}^{\dag} & 0 \\
0 & 0 & 0 & {\mathcal{D}}_{2}^{\dag}{\mathcal{D}}_{2}  \\
\end{array} \right)
.\end{equation}
In addition, the Witten parity operator is:
\begin{equation}\label{wparityopera}
\mathcal{W}_H= \left ( \begin{array}{cccc}
  1 & 0 & 0 & 0 \\
  0 & -1 & 0 & 0 \\
0 & 0 & 1 & 0 \\
0 & 0 & 0 & -1  \\
\end{array} \right)
.\end{equation}
The operators (\ref{connectirtyrtons}), (\ref{connections1dtr}) and (\ref{wparityopera}), satisfy the following relations:
\begin{equation}\label{mousikisimagne}
\{ {\mathcal{Q}}_{H},{\mathcal{Q}}_{H}^{\dag}\}=H_{H},{\,}{\,}{\mathcal{Q}}_{H}^2=0,{\,}{\,}{{\mathcal{Q}}_{H}^{\dag}}^2=0,{\,}{\,}\{{\mathcal{Q}}_{H},\mathcal{W}_H\}=0,{\,}{\,}\mathcal{W}_H^2=I,{\,}{\,}[\mathcal{W}_H,H_{H}]=0
.\end{equation}
Similar to the representation (\ref{connectirtyrtons}) for the supercharges, we can obtain equivalent higher dimensional representations for the combined $N=2$, $d=1$ algebra, by making the following replacements:
\begin{equation}\label{setoftransformations}
\mathrm{Set}{\,}{\,}{\,}A:{\,}
\begin{array}{c}
 {\mathcal{D}}_{1}\rightarrow {\mathcal{D}}_{1}^{\dag} \\
  {\mathcal{D}}_{2}^{\dag}\rightarrow {\mathcal{D}}_{2} \\
\end{array},{\,}{\,}{\,}\mathrm{Set}{\,}{\,}{\,}B:{\,}
\begin{array}{c}
 {\mathcal{D}}_{1}\rightarrow {\mathcal{D}}_{2}^{\dag} \\
  {\mathcal{D}}_{2}^{\dag}\rightarrow {\mathcal{D}}_{1} \\
\end{array},{\,}{\,}{\,}\mathrm{Set}{\,}{\,}{\,}C:{\,}
\begin{array}{c}
 {\mathcal{D}}_{1}\rightarrow {\mathcal{D}}_{2} \\
  {\mathcal{D}}_{2}^{\dag}\rightarrow {\mathcal{D}}_{1}^{\dag} \\
\end{array}
.\end{equation}
We omit for brevity the details of each representation presented in the relation (\ref{setoftransformations}), since these are similar to relation (\ref{connectirtyrtons}).

\subsection{A $q$-deformed Extension of each $N=2$ SUSY QM Algebra}

Consider the algebra of relation (\ref{relationsforsusy}) and the supercharges and Hamiltonian of relation (\ref{s7}). By modifying the Hamiltonian $\mathcal{H}_1$ to have the following form:
\begin{equation}\label{hffh}
\mathcal{H}_{1}=\bigg{(}\begin{array}{ccc}
 q\mathcal{D}_{1}\mathcal{D}_{1}^{\dag} & 0 \\
  0 & q^{-1}\mathcal{D}_{1}^{\dag}\mathcal{D}_{1}  \\
\end{array}\bigg{)}
\end{equation}
we automatically modify the algebraic commutation and anti-commutation relations of relation (\ref{relationsforsusy}), which no longer hold true. However, there is a new algebra, which is known as a $q$-deformed extension \cite{qdef} of the SUSY QM algebra (\ref{relationsforsusy}), for which new commutation and anti-commutation relations hold true. These are the following relations:
\begin{equation}\label{qde}
\{\mathcal{Q}_{1},\mathcal{Q}^{\dag}_{1}\}_q=\mathcal{H}_{1}{\,}{\,},\{\mathcal{Q}_{1},\mathcal{Q}_{1}\}_q=0,{\,}{\,}\{\mathcal{Q}_{1}^{\dag},\mathcal{Q}_{1}^{\dag}\}_q=0
\end{equation}
and additionally,
\begin{equation}\label{qdddfre}
[\mathcal{H}_1,\mathcal{Q}_{1}]_q=[\mathcal{H}_1,\mathcal{Q}_{1}^{\dag}]_q=0
\end{equation}
In relations (\ref{qde}) and (\ref{qdddfre}) we used the $q$-deformed version of the commutator and the anti-commutator which are defined as follows:
\begin{equation}\label{comanticomdef}
[X,Y]_q=qXY-q^{-1}YX,{\,}{\,}{\,}\{X,Y\}_q=qXY+q^{-1}YX,
\end{equation}
and satisfy:
\begin{equation}\label{dshaydsd}
[Y,X]_q=-[X,Y]_{q^{-1}},{\,}{\,}{\,}\{Y,X\}_q=\{X,Y\}_{q^{-1}}
\end{equation}
Hence, the simple $N=2$ SUSY QM algebra of the system with Hamiltonian $\mathcal{H}_1$ may be trivially extended to a $q$-deformed one. Of course, the same applies to the other $N=2$ SUSY QM algebra. After presenting this additional algebraic structure, we shall investigate the possibility that the two $N=2$ SUSY QM algebras combine to give an enhanced supersymmetric structure. This is the topic of section 3.

\subsection{Witten Index, Addition of non-renormalizable mass terms and Static Background Electromagnetic Fields}

One question we shall address in this section, is whether the Witten index remains invariant in the case we add some non-renormalizable mass terms for the fermions and in addition if the index remains invariant in the case we turn on static background electromagnetic fields. In the case of electromagnetic fields, this problem is quite interesting and a study on the correspondence of superconducting defects in the presence of electromagnetic fields was done in \cite{witten2}. A solid criterion in order to investigate the compact perturbations of the Witten index problem, is a theorem that deals with compact perturbations of the index of Fredholm operators. Let us take into account the system corresponding to equations (\ref{finaleqns}) and to the operator (\ref{susyqmrn567m}), which as we evinced is Fredholm. Let $C$ an odd compact symmetric matrix \cite{thaller}
\begin{equation}\label{susyqmrnmassive}
\mathcal{C}=\left(%
\begin{array}{cc}
 0 & \mathcal{C}_1
 \\ \mathcal{C}_2 & 0\\
\end{array}%
\right)
\end{equation}
Consider the operator $\mathcal{D}_{p}=\mathcal{D}_{1}+\mathcal{C}$, which is actually a compact perturbation of the operator $\mathcal{D}_1$. Since the operator $\mathrm{tr}\mathcal{W}e^{-t(\mathcal{D}_1+\mathcal{C})^2}$ is trace class (which is true since compact perturbations of Fredholm operators are also Fredholm operators and hence trace-class), the following theorem holds (see for example \cite{thaller} page 168, Theorem 5.28),
\begin{equation}\label{indperturbatrn1}
\mathrm{ind}\mathcal{D}_p=\mathrm{ind}(\mathcal{D}_1+\mathcal{C})=\mathrm{ind}\mathcal{D}_1
,\end{equation}
with $C$ the symmetric odd operator of the form (\ref{susyqmrnmassive}). Therefore, by virtue of the theorem (\ref{indperturbatrn1}) and owing to relation (\ref{ker1}), the Witten index of the SUSY QM system corresponding to operator $\mathcal{D}_1$, is invariant under compact perturbations of the operator $\mathcal{D}_1$. As we shall demonstrate, only the case corresponding to turning on a background fast decaying or constant static magnetic field produces compact perturbations of the operator $\mathcal{D}_1$. In order to see this, we shall present in detail all the cases listed in the beginning of this section.

Consider first the addition of non-renormalizable Majorana and Dirac mass terms. The Lagrangian terms that correspond to these terms are of the form,
\begin{equation}\label{dirlag}
\mathcal{L}_D=-\frac{1}{M}y_D(\psi_L^T\sigma_2\Phi)C^{-1}(\Phi^T\sigma_2\psi_L^T)+\mathrm{h.c.}
\end{equation}
for the Dirac case and 
\begin{equation}\label{dirlagmag}
\mathcal{L}_D=-\frac{\eta^2y_M}{2M}\psi_L^TC^{-1}\psi_L^T+\mathrm{h.c.}
\end{equation}
for the left-handed Majorana mass term. In the above relations, $M$ is the mass scale corresponding to new physics, characteristic of the energy order at which the Standard Model is modified. In addition, $y_D$ and $y_M$ is the Yukawa couplings corresponding to the Dirac and Majorana mass terms. Under the addition of these two terms, the operator $\mathcal{D}_1$ is modified as follows:
\begin{equation}\label{noodd}
\mathcal{D}_1+\mathcal{F}
\end{equation}
with $\mathcal{F}$ an even matrix, which depends on the new mass terms (\ref{dirlag}) and (\ref{dirlagmag}) (we omit the details of it for simplicity). Since the matrix $\mathcal{F}$ is even, the theorem (\ref{indperturbatrn1}) does not apply and hence, the Witten index does not remain invariant. The same applies in the case we turn on a static electric field. Let us turn our interest in the case we turn on a magnetic field $\mathcal{B}=(\mathcal{B}_x,\mathcal{B}_y,\mathcal{B}_z)$. The first two components modify equations (\ref{finaleqns}) in such a way so that operator $\mathcal{D}_1$ is modified in the same way as in equation (\ref{noodd}), that is, operator $\mathcal{D}_1$ is perturbed by the addition of an even compact operator (recall we consider static fast decaying magnetic fields piercing the domain wall). Only the last component of $\mathcal{B}$, generates compact and odd perturbations of operator $\mathcal{D}_1$. We shall now present in detail this case. Consider a electromagnetic field of the form $\mathcal{A}^{\mu}=(0,0,0,\mathcal{B}_z)$, with $\mathcal{B}_z$ a constant or fast decaying magnetic field (constant or fast decaying in order the operators are compact, as we shall see). Including the background magnetic field interaction with the fermions, the equations (\ref{finaleqns}) can be written as follows:
\begin{align}\label{finaleqnsfinalchap}
&(i\sigma^3\partial_0-\sigma^2\partial_1+\sigma^1\partial_2+\mathcal{M})\Psi_A+(i\partial_3+\mathcal{B}_z)\Psi_B=0\\ \notag &
(i\sigma^3\partial_0-\sigma^2\partial_1+\sigma^1\partial_2-\mathcal{M})\Psi_B+(-i\partial_3-\mathcal{B}_z)\Psi_A=0
\end{align}
and consequently the operator $\mathcal{D}_1$ takes an additional contribution coming from the magnetic field $\mathcal{B}_z$. Particularly, it can be written as follows:
\begin{equation}\label{chaniaaddop}
\mathcal{D}_1'=\mathcal{D}_1+\mathcal{C}_z,
\end{equation}
with $\mathcal{C}_z$ being equal to:
\begin{equation}\label{susyqmrnmafgreh}
\mathcal{C}_z=\left(%
\begin{array}{cc}
 0 & \mathcal{B}_z
 \\ -\mathcal{B}_z & 0\\
\end{array}%
\right)
\end{equation}
The operator $\mathcal{C}_z$ is odd and in order theorem (\ref{indperturbatrn1}) holds true, we require that the operator $\mathcal{B}_z$ to be compact. This can be true when $\mathcal{B}_z$ is a finite constant, or if it decays fast. Now the functional dependence of $\mathcal{B}_z$ is of no mathematical importance, but the physics underlying the choice of $\mathcal{B}_z$ can be interesting, since it has to do with a magnetic field near or on a domain wall. We shall not be interested in the cosmological outcomes of this, but we are mostly interested for the mathematical implications of compact perturbations of the system. Hence if the operator $\mathcal{B}_z$ is compact, theorem (\ref{indperturbatrn1}) holds true and therefore the Witten index of the system is invariant, that is:
\begin{equation}\label{fhjhggfgy60seconds}
\mathrm{ind}\mathcal{D}_1'=\mathrm{ind}(\mathcal{D}_1+\mathcal{C}_z)=\mathrm{ind}\mathcal{D}_1
\end{equation} 
This means that the net number of scattered fermionic modes of the system remains unaltered. This is potentially interesting, since the net number of scattered solutions off a domain wall, remains invariant under the influence of a magnetic field with a component parallel to the $z$-direction of the domain wall (recall the domain wall extends to the $xz$-plane). 

Before closing, let us discuss a rather interesting coincidence. It is known from the literature \cite{iapon1,iapon2} that the localized fermionic zero modes on the domain wall produce spontaneously a magnetic field on the domain wall \cite{iapon1,iapon2}. Although there is a controversy on whether this magnetic field extends away from the wall (cosmological distances) \cite{iapon1}, or it has finite range \cite{iapon2}, the existence of the magnetic field on the wall is due to the massless fermions localized on the wall \cite{iapon1,iapon2}. Thus the problem at hand, that is, the calculation of the index under constant magnetic field perturbations, is useful and that was the main motivation to study magnetic field perturbations. It is rather intriguing that the localized zero modes on the domain wall, that also have a rich SUSY QM structure, affect the scattered zero modes through magnetic field perturbations, without implying of course any direct correlation between the two SUSY QM structures.

\section{$N=4$ Extended Supersymmetric Quantum Mechanics Algebras in the Scattering States}

Having found the two supersymmetric quantum mechanics algebras underlying the scattered fermionic states, it is natural to ask if there is an enhanced supersymmetric structure underlying the system. As we shall demonstrate in this section, the answer lies in the affirmative. Particularly, the enhanced supersymmetric structure is a $N=4$ SUSY QM with central charge. In order to see this, we compute the following commutation and anti-commutation relations:
\begin{align}\label{commutatorsanticomm}
&\{{{\mathcal{Q}'}_{2}},{{\mathcal{Q}}_{2}}^{\dag}\}=2\mathcal{H},{\,}\{{{\mathcal{Q}}_{1}},{{\mathcal{Q}}_{1}}^{\dag}\}=2\mathcal{H},{\,}\{{{\mathcal{Q}}_{1}},{{\mathcal{Q}}_{1}}\}=0,{\,}\{{{\mathcal{Q}}_{2}},{{\mathcal{Q}}_{2}}\}=0,{\,}{\,}\\
\notag & \{{{\mathcal{Q}}_{1}},{{\mathcal{Q}}_{2}}^{\dag}\}=\mathcal{Z},{\,}\{{{\mathcal{Q}}_{2}},{{\mathcal{Q}}_{1}}^{\dag}\}=\mathcal{Z},{\,}\\ \notag
&\{{{\mathcal{Q}}_{2}}^{\dag},{{\mathcal{Q}}_{2}}^{\dag}\}=0,\{{{\mathcal{Q}}_{1}}^{\dag},{{\mathcal{Q}}_{1}}^{\dag}\}=0,{\,}\{{{\mathcal{Q}}_{1}}^{\dag},{{\mathcal{Q}}_{2}}^{\dag}\}=0,{\,}\{{{\mathcal{Q}}_{1}},{{\mathcal{Q}}_{2}}\}=0{\,}\\
\notag
&[{{\mathcal{Q}}_{2}},{{\mathcal{Q}}_{1}}]=0,[{{\mathcal{Q}}_{1}}^{\dag},{{\mathcal{Q}}_{2}}^{\dag}]=0,{\,}[{{\mathcal{Q}}_{2}},{{\mathcal{Q}}_{2}}]=0,{\,}[{{\mathcal{Q}}_{2}}^{\dag},{{\mathcal{Q}}_{2}}^{\dag}]=0,{\,}\\
\notag &
[{\mathcal{H}}_{2},{{\mathcal{Q}}_{2}}]=0,{\,}[{\mathcal{H}}_{2},{{\mathcal{Q}}_{2}}^{\dag}]=0,{\,}[\mathcal{H}_{1},{{\mathcal{Q}}_{1}}^{\dag}]=0,{\,}[\mathcal{H}_{1},{{\mathcal{Q}}_{1}}]=0,{\,}
\end{align}
with $\mathcal{Z}$:
\begin{equation}\label{zcentralcharge}
\mathcal{Z}=2\mathcal{H}_{1}=2{\mathcal{H}}_{2}
\end{equation}
The operator $\mathcal{Z}$ commutes with all the operators corresponding to the $N=2$ algebras, namely the
supercharges ${{\mathcal{Q}}_{1}},{{\mathcal{Q}}_{2}}$, their conjugates ${{\mathcal{Q}}_{1}}^{\dag},{{\mathcal{Q}}_{2}}^{\dag}$ and
finally the Hamiltonians, $\mathcal{H}=\mathcal{H}_{1}={\mathcal{H}}_{2}$.
The above relations
(\ref{commutatorsanticomm}) constitute a central charge extended $N=4$ supersymmetric quantum mechanics
algebra with central charge $\mathcal{Z}$, which is the quantum Hamiltonian of each $N=2$, $d=1$ subsystem. It would be convenient to recall that the $N=4$ SUSY QM algebra with central charge is described by the following relations:
\begin{align}\label{n4algbe}
&\{Q_i,Q_j^{\dag}\}=2\delta_i^jH+Z_{ij},{\,}{\,}i=1,2 \\ \notag &
\{Q_i,Q_j\}=0,{\,}{\,}\{Q_i^{\dag},Q_j^{\dag}\}=0
\end{align}
In our case, the SUSY QM algebra of relation (\ref{commutatorsanticomm}) has two central charges, namely $Z_{12}$ and $Z_{21}$ which
are equal, that is:
\begin{equation}\label{fdjjf}
Z_{12}=Z_{21}=\mathcal{Z}
\end{equation}
The other two possible supercharges are equal to zero, due to the structure of the operators $\mathcal{D}_1$ and $\mathcal{D}_2$, that is:
\begin{equation}\label{othertwosupercharges}
Z_{11}=Z_{22}=0
\end{equation}
\noindent Obviously, such an $N=4$ SUSY QM structure is particularly interesting, since a similar structure underlies the localized fermions on the domain wall, as evinced in \cite{lazaridesvasiko}.

\noindent The $N=4$ supersymmetric algebra is quite frequently met in string theory contexts, since extended (with $N=4,6...$) supersymmetric quantum mechanical models are models that result by dimensionally reducing $N=2$ and $N=1$ Super-Yang Mills theories to one temporal dimension. Furthermore, extended supersymmetries are directly related to super-extensions of
integrable models (Calogero-Moser
systems), and also to super-extensions of Landau-type models \cite{ivanov}.

\section{Implications of the SUSY QM Algebras on the Hilbert Space of the Fermionic Scattering States}

\subsection{A Global $U(1)_A\times U(1)_B$ of the Hilbert Space of Quantum States}

The $N=2$ SUSY QM algebra equips the Hilbert space of the fermionic quantum states with many additional algebraic and geometric structures which we now describe in detail. An explicit structure that can be easily seen, is the existence of a global gauge symmetry. Particularly, due to the $N=2$ SUSY QM, there exists a global $U(1)$ symmetry in the fermionic states. Since the system has two independent $N=2$, $d=1$ symmetries, the total vector space of fermionic states is equipped with a product of two such $U(1)$ symmetries. For simplicity we focus on the system with supercharges and Hamiltonian those of relation (\ref{relationsforsusy}). The $N=2$, $d=1$ SUSY QM algebra of relation (\ref{relationsforsusy}) is
invariant under the global $U(1)$ transformations:
\begin{align}\label{transformationu1}
& {\mathcal{Q}}_{1}^{'}=e^{-ia}{\mathcal{Q}}_{1}, 
\\ \notag &{\mathcal{Q}}^{'\dag}_{1}=e^{ia}{\mathcal{Q}}^{\dag}_{1}
.\end{align}
Therefore, the quantum system is actually invariant under an $R$-symmetry which is of the
form of a global-$U(1)$. Moreover, the Hamiltonian of the SUSY QM algebra $\mathcal{H}_1$ is invariant under the $U(1)$-transformation, that is, $\mathcal{H}_1'=\mathcal{H}_1$. The global $U(1)$ symmetry we just described, has a direct impact on the Hilbert space of the quantum system, which is formed from the fermionic states scattered off the domain wall. Particularly, it imposes certain transformation properties on the fermionic states.  Let us see this in detail. Recall that the total quantum Hilbert space of the system $\mathcal{H}$, is $Z_2$-graded. We denote $\psi^{+}_{M}$ and
$\psi^{-}_{M}$, the Hilbert states corresponding to the
spaces $\mathcal{H}^{+}_{M}$ and $\mathcal{H}^{-}_{M}$ respectively. The global $U(1)$ symmetry of the quantum algebra (\ref{transformationu1}) implies the following transformations on the Hilbert quantum states:
\begin{align}
&\psi^{'+}_{M}=e^{-i\beta_{+}}\psi^{+}_{M},
\\ \notag &
\psi^{'-}_{M}=e^{-i\beta_{-}}\psi^{-}_{M}
.\end{align}
with $\beta_{+}$ and $\beta_{-}$ global
parameters defined in the following way:
\begin{equation}\label{fhffn}
a=\beta_{+}-\beta_{-}
\end{equation}
Hence, the fermionic SUSY QM quantum system of relation (\ref{relationsforsusy}) possesses a global U(1)-symmetry. The same arguments hold true for the other SUSY QM quantum system described by relations (\ref{sdff7}), so an additional global $U(1)$ symmetry implies similar transformation properties for the fermionic quantum states of the second SUSY QM system. Denoting $U(1)_A$ the global gauge symmetry of the system (\ref{relationsforsusy}) and with $U(1)_B$ the global gauge symmetry of the system described by relation (\ref{sdff7}), the total global symmetry of the scattered fermions is $\mathcal{G}_S$, which is:
\begin{equation}\label{alexanderthegreat}
\mathcal{G}_S=U(1)_A\times U(1)_B
\end{equation}

\noindent In reference to scattered states, there exist bound states in the spectrum which are important too, since these are the localized states on the domain wall. As it is known, discrete symmetries are inherent to quantum system with fermionic condensates, so the existence of a global symmetry and the study of its breaking is particularly important. Noticeable is the fact that an explicit breaking could make us assume that a supercharge acquires a constant vacuum expectation value. We defer this study to a future work.

\subsection{Some Local Geometric Implications of the SUSY QM Algebra on the Spacetime Fibre Bundle Structure of the Scattered Fermions}

The existence of an $N=2$ SUSY QM algebra in the scattered fermions system, has some local geometric implications on the fibre bundle structure of the spacetime $M$, on which the fermions are defined. As we now demonstrate, owing to the $N=2$ SUSY QM algebra, the spacetime manifold $M$ is locally  a supermanifold. Particularly, the supercharges of the SUSY QM algebra have a mathematical meaning on the supermanifold, with the supercharge of the SUSY QM algebra being the local superconnection on this supermanifold, and the square of the supercharge being the corresponding curvature. For the details on the mathematical issues we shall present see \cite{graded1}. For convenience we shall focus on the first $N=2$, $d=1$ algebra with supercharge $\mathcal{Q}_1$.

\noindent The scattered fermions off the domain walls, in the spacetime $M$, are sections of the $U(1)$-twisted fibre bundle $P\times S \otimes U(1)$, with $S$ the reducible representation of the Spin group $Spin(4)$, and $P$ is the double cover of the principal $SO(4)$ bundle on the tangent manifold $TM$. Recall that on the Hilbert space of the fermionic states there exists a $Z_2$ grading. In general, a $Z_2$ grading on a vector space $E$, is performed by decomposing the vector space as follows:
\begin{equation}\label{z2grad}
E=E_+\oplus E_{-}
\end{equation}
Moreover, a $Z_2$-grading of an algebra $A$ to even and odd elements, $A=A_+\oplus A_{-}$, is performed in such a way so that the following relations are satisfied: 
\begin{equation}\label{amodule}
A_+\cdot E_+\subset E_+,{\,}{\,}A_+\cdot E_-\subset
E_-,{\,}{\,}A_-\cdot E_+\subset E_-,{\,}{\,}A_-\cdot E_-\subset
E_+,
\end{equation}
If these relations hold true, the algebra $A$ is called a $Z_2$-graded algebra. We denote with $\mathrm{End} (E)$, the set of endomorphisms of $E$. The Witten parity operator of the SUSY QM algebra belong to the set of endomorphisms $\mathrm{End} (E)$. The involution $\mathcal{W}$ acts on the vectors of the vector space $E$, as follows:
\begin{equation}\label{befoend}
\mathcal{W}(a+b)=a-b,{\,}{\,}{\,}\forall{\,}{\,}a{\,}\in{\,}E_+,{\,}{\,}\mathrm{and}{\,}{\,}\forall{\,}{\,}b{\,}\in{\,}E_-.
\end{equation} 
The Witten parity operator $\mathcal{W}$ (which is an involution operator) plays a crucial role since it provides the algebra $\mathrm{End} (E)$ with a $Z_2$-grading.

Let us see what is the algebraic impact of this $Z_2$-grading on the $N=2$ SUSY QM quantum Hilbert space $\mathcal{H}$ of the scattered fermions.  The involution $\mathcal{W}$, generates the $Z_2$ graded vector space
$\mathcal{H}=\mathcal{H}^+\oplus \mathcal{H}^-$. The subspace
$\mathcal{H}^+$ contains $\mathcal{W}$-even vectors while $\mathcal{H}^-$, $\mathcal{W}$-odd vectors.
Therefore, we can define an additional $Z_2$-graded algebra $\mathcal{A}$, with
$\mathcal{A}=\mathcal{A}^+\oplus \mathcal{A}^-$, on the manifold $M$. In our case, the algebra $\mathcal{A}$ is a total rank two sheaf of
$Z_2$-graded commutative $R$-algebras. Therefore, $M$
becomes a graded manifold $(M,\mathcal{A})$. However, this extra algebraic structure does not render the manifold $M$ a supermanifold, at least globally.
\noindent The endomorphism $\mathcal{W}$ which acts as,
\begin{equation}\label{dsdf}
\mathcal{W}:\mathcal{H}\rightarrow \mathcal{H}
\end{equation}
is a non-trivial element of the sheaf $\mathcal{A}$, and therefore
\begin{equation}\label{polyvroxa}
\mathrm{End}(\mathcal{H})\subseteq \mathcal{A}
\end{equation}
The sheaf $\mathcal{A}$ is the structure sheaf of the graded manifold
$(M,\mathcal{A})$, and the manifold $M$ is the body of
$(M,\mathcal{A})$. The structure sheaf $\mathcal{A}$ is locally isomorphic to the sheaf
$C^{\infty}(U)\otimes\wedge R^m$ of the exterior affine vector bundle
$\wedge \mathcal{H_E}^*=U\times \wedge R^m$. The affine vector bundle $\mathcal{H_E}$
has as fiber the space $\mathcal{H}$ and $U$ denotes an arbitrary open set of the manifold $M$. The structure sheaf $\mathcal{A}=C^{\infty}(U)\otimes\wedge \mathcal{H}$, is
isomorphic to the sheaf of sections of the exterior vector bundle
$\wedge \mathcal{H_E}^*=R\oplus
(\oplus^{m}_{k=1}\wedge^k)\mathcal{H_E}^*$. The sheaf structure we just described has some local geometric implications on the manifold $M$. The sections of the fibre bundle $TM^*\otimes\mathcal{H}$ are the sections of the fermionic bundle $P\times S \otimes U(1)$. A local superconnection, denoted $\mathcal{S}$, is an 1-form which takes values in $\mathrm{End} (E)$. Stated differently, a local superconnection is a section of  $TM^*\otimes \wedge \mathcal{H_E}^*\otimes
\mathcal{H_E}$. The corresponding curvature of the superconnection, denoted $\mathcal{C}$, is an $\mathrm{End} (E)$-valued 2-form on $M$, with:
\begin{equation}\label{}
\mathcal{C}=\mathcal{S}^2
\end{equation}
Hence, the superconnection is a section of the fibre bundle $TM^*\otimes \mathrm{End} (E)^{\mathrm{odd}}$, at least locally on $M$. The latter contains the odd elements of $\mathrm{End} (E)$. Thereby, at an infinitesimally small open neighborhood of a point $x$ $\in$ $M$, the supercharge of the SUSY QM algebra is identified with the superconnection, that is $\mathcal{S}={\mathcal{Q}_1}$. Consequently, the curvature of the supermanifold is locally, 
\begin{equation}\label{poloyvroxei}
\mathcal{C}={\mathcal{Q}}^2_1
\end{equation}

\noindent In conclusion, the $N=2$ SUSY QM structure locally makes the manifold $M$ a supermanifold, with the supercharge ${\mathcal{Q}_1}$ being its superconnection and the square of the supercharge ${\mathcal{Q}}^2_1$ being the curvature of the supermanifold locally. However, globally the manifold $M$ is a graded manifold $(M,\mathcal{A})$ with body $M$ and structure sheaf $\mathcal{A}$. 

\subsection{SUSY QM and Global Supersymmetry}

As we demonstrated in the previous section, the manifold $M$ is locally a supermanifold. A natural question that springs to mind is whether there is any connection of the SUSY QM algebra with a global spacetime supersymmetry. The answer is no.

When studying supersymmetric algebras in various dimensions we have to bear in mind that the usual spacetime supersymmetric algebra (which is the graded super-Poincare algebra in four dimensions) is four dimensional while the SUSY QM algebra is one dimensional. Moreover, spacetime supersymmetry in $d>1$ dimensions and
SUSY QM, which is an $d=1$ supersymmetry, are in principle different issues, with the only possible correlation being the fact that extended (with $N = 4, 6...$) SUSY QM models are obtained by the dimensional reduction of $N = 2$ and $N = 1$
Super-Yang Mills models to one dimension  \cite{ivanov}. However, the complex supercharges of $N = 2$, $d=1$ SUSY QM are not related in any way to the generators of spacetime supersymmetry and hence, SUSY QM does not relate fermions and bosons. Note that by saying fermions and bosons we mean the representations of the super-Poincare graded Lie algebra in four dimensions. Nevertheless, the SUSY QM supercharges render the Hilbert space of quantum states a $Z_2$ graded vector space. In addition, these supercharges generate transformations between the Witten parity eigenstates, a fact that explains why the manifold $M$ has globally the structure of a graded manifold and not that of a supermanifold.

\section*{Concluding Remarks}

In this paper we studied the quantum system of scattered fermions off a domain wall in the thin-wall approximation. As we demonstrated, the system possesses two one dimensional $N=2$ SUSY QM algebras and therefore the Hilbert space of the fermionic quantum states has an additional algebraic structure, which is that of a global graded manifold. In addition, the two $N=2$ algebras combine to form an one dimensional $N=4$ SUSY QM algebra with non-trivial central charge. However, there is no general rule that governs this result, that is, finding extended supersymmetric structures when many $N=2$ are present in the initial system. In fact, there are cases in which the initial system does have $N=1$ global supersymmetry and the quantum Hilbert space of the states has only one $N=2$ SUSY QM algebra and not an extended one, as someone would suspect (see for example reference \cite{oiko1} and references therein).

Moreover, we studied how the Witten index of each $N=2$ SUSY QM algebra responds under even and odd compact perturbations of the operators involved. The even perturbations are generated when non-renormalizable Yukawa mass terms for the fermions are taken into account. In addition, even perturbations are generated when background static magnetic fields are taken into account and specifically when the $(\mathcal{B}_x,\mathcal{B}_y)$ components are considered. As we saw, the Witten index is not invariant under these even perturbations. The same applies for static electric fields. On the contrary, odd perturbations leave the Witten index invariant. In our case, odd perturbations can be generated when the background magnetic field has only one non-zero component, namely the $B_z$ component. Such a result is kind of interesting, since magnetic fields are associated to the domain walls, owing to the existence of localized fermionic zero modes.

In principle, the total complexity of a general problem may be reduced if we develop techniques to find all possible internal symmetries. In addition, if instead of symmetries we discover any regularity or repeating underlying pattern, we accomplish even more deep insight to the problem at hand. In reference \cite{rossi} was point out that the existence of the fermionic zero modes can be associated to the presence of a hidden underlying symmetry that the system possesses. This is somehow different in spirit from problems of localized fermions on higher dimensional defects like branes (for an important stream of papers on this, see for example \cite{liu} and references therein). This symmetry was evinced to be some kind of supersymmetry \cite{rossi}. Hence, it is very intriguing the fact that also the scattered states of the fermionic system at hand are directly associated to $N=2$ and $N=4$ one dimensional supersymmetries. These SUSY QM algebras can be remnants of this hidden supersymmetry in some way. This study is rather interesting and we hope to address these issues in the future.

Finally, since the SUSY QM algebra is an attribute of both Majorana and Dirac fermions, this algebra could provide information and probably this information may have its imprint on the energy spectrum of the fermions. Particularly interesting case of fermions scattered off domain walls are the neutrinos. Moreover, in reference to neutrinos, since the nature of the neutrino (Dirac or Majorana) is yet to be understood-revealed, studies of such interactions may provide us with additional information. In addition, as we saw, since the operators we used in this paper are compact, a compact odd perturbation on the neutrino mass does not change the Witten index, and thus does not change the net number of scattered solutions.

\section*{Acknowledgments} Financial support by the Research
Committee of the Technological Education Institute of Central Macedonia,
under grant SAT/ME/190314-10/7, is gratefully acknowledged. V. O  would like to thank Prof. I. Antoniadis for a brief but very helpful discussion on central charges.

\end{document}